\documentclass[epj]{svjour}
\usepackage{amsfonts}
\newcommand{\fant}[1]{\phantom{#1}}
\newcommand{\be}{\begin{equation}}
\newcommand{\ee}{\end{equation}}
\newcommand{\wdg}{\wedge}

\begin{document}
\title{Modified Einstein-Gauss-Bonnet gravity: Riemann-Cartan and Pseudo-Riemannian cases}
\author{Hatice \"Ozer\inst{1}\thanks{\emph{\email{hatice.ozer@istanbul.edu.tr}} } 
\and Ahmet Baykal\inst{2}\thanks{\emph{\email{abaykal@nigde.edu.tr,failure2communicate@gmail.com}} } \and \"Ozg\"ur Delice\inst{3}\thanks{\emph{\email{ozgur.delice@marmara.edu.tr}} }
}                    
%
%
\institute{Department of Physics, Faculty of  Sciences,  \.Istanbul University, 34134 \.Istanbul, Turkey \and Department of Physics, Faculty of Arts and Sciences, 
Ni\u gde University,  51240 Ni\u gde, Turkey \and Department of Physics, Faculty of Arts and Sciences, Marmara University, 34722  Istanbul, Turkey 
}
%

\abstract{
A modified Einstein-Gauss-Bonnet gravity in four dimensions where the quadratic Gauss-Bonnet  term is coupled to  a scalar field is considered. The field equations of the model are obtained by variational methods by making use of the constrained-first order formalism covering  both pseudo-Riemannian and non-Riemannian cases.  
In the pseudo-Riemannian case, the Lagrange multiplier forms, which impose the vanishing torsion constraint, are eliminated in favor of the remaining fields and  the resulting metric field equations are expressed in terms of the double-dual curvature 2-form.  In the  non-Riemannian case with torsion, the  field equations  are expressed in terms of the pseudo-Riemannian quantities by a perturbative scheme valid for a weak coupling constant. It is shown that, for both cases, the model admits a maximally symmetric de-Sitter solution  with  nontrivial scalar field. Minimal coupling of a Dirac spinor to the Gauss-Bonnet modified gravity is also discussed briefly.
} 
\maketitle

\section{Introduction}

Einstein's general relativity (GR) theory, of which we are celebrating its 100th anniversary, passes all tests in the weak  gravity regime. 
However, research on  alternative theories does not seem to come to an end. There are various reasons for investigating alternative theories. On the observational side, the observations on the galactic scale such as  the galaxy rotation curves are explained by introducing the dark matter whereas  on the other hand, on the cosmological scale, the late time acceleration of the universe \cite{darken1,darken2,darken3} concluded from the observations of  type-Ia supernovae leads to the concept of dark energy, which behaves like a cosmological constant. With these dark components included,  Einstein's theory perfectly explains large scale behaviour of the universe with the so called $\Lambda$-CDM model. Although these dark components, especially  dark matter, are awaiting their exploration  by some high energy experiments, the idea that this large scale behaviour can be explained by modifying GR becomes plausible.  Moreover, on theoretical grounds, several different ideas propose that the modifications of GR  be taken into account. For example, in the string theory, the inclusion of higher order curvature components to the  usual Einstein-Hilbert action, such as the Gauss-Bonnet  term \cite{chern,lovelock1,lovelock2},  is implied by high energy corrections \cite{boulware-deser}.  There is enormous literature on modified gravity theories,  hence we  refer the reader to the latest reviews  \cite{sotiruou-faraoni,nojiri-odintsov-rev,capozziello,padilla} for further motivations and  for the latest developments on this subject.

The Gauss-Bonnet (GB) Lagrangian, the second term of the Euler-Poincare forms \cite{chern}, is a natural generalization \cite{lovelock1,lovelock2} of Einstein's theory into higher dimensions where the first term is the Einstein-Hilbert term itself.  Remarkable properties of GR are also present in the Lagrangians involving these forms such as being ghost free  and having at most second order in the metric. However, in four spacetime dimensions, GB Lagrangian does not contribute to the field equations since it becomes a topological invariant. However, if coupled to a scalar field, it can contribute to the equations. Another way of  doing this is by considering a function of GB term denoted as $f(\mathcal{G})$ theories.
These modified Gauss-Bonnet gravities are well known and their properties and some exact solutions in four or higher dimensions can be found in the literature 
\cite{boulware-deser,boulware-deser-2,dereli-tucker-effective-string-models,wetterich,ozkurt-dereli,gasperini-giovannini,ozkurt1,ozkurt2,nojiri-odintsov-sasaki,nojiri-odintsov,Cognola,Koivisto-mota1,Koivisto-mota2,cognola-elizalde-nojiri,bamba,sotiruou,gurses,mielke,saridakis,uddin,capozziello-elizalde,defelice,h-j-schmidt,rodrigues}.   

Yet another minimal way to modify  GR is to modify the differential geometric setting by considering a more general non-Riemannian connection involving a nonvanishing torsion, as in the Einstein-Cartan (EC) theory \cite{cartan1,cartan2,cartan3,sciama,kibble}. Notice that in EC theory, if there are no spinor matter fields having torsional contribution or in vacuum, the field equations will be exactly the same as in GR. On the other hand,  if we consider general scalar-tensor theories, such as Brans-Dicke theory, the nonminimal coupled scalar field generates torsion \cite{dereli-tucker-torsion}.  It is well-known that the torsion in this particular case is algebraic and can be eliminated in favor  of the scalar field and then the terms arising from torsion can be expressed  in terms of pseudo-Riemannian  quantities. Hence, the torsional contribution of the scalar field  can be encoded into a pseudo-Riemannian theory as an additional matter-energy source \cite{dereli-tucker-torsion},  which leads to additional scalar interaction terms compared to the  BD theory in the pseudo-Riemannian context. This point of view is also extended to non-Riemannian $f(R)$ theories in a similar manner, see, for example, \cite{padilla,baykal}.

The gravitational theories involving bare Gauss-Bonnet terms with torsion were discussed before \cite{troncoso} and some exact solutions were obtained \cite{canfora}. Therefore,  it might be worthwhile to consider a similar approach for theories involving some scalar field now coupling to powers of curvature in a GB Lagrangian 4-form.

The paper is organized as follows. After introducing the notation for the geometrical quantities  in sect. 2,  we consider  the first order constrained variation technique to the modified GB theory with a non-minimally coupled scalar field, obtain the Lagrange multiplier form and express the field equations in a compact way by using the exterior algebra of differential forms in sect. 3. In the following section, we discuss the non-Riemannian theory with nonvanishing  torsion following from the same Lagrangian density considered. After presenting the field equations involving a propagating torsion, we focus on the independent connection equations. In order to reduce the complicated field equations with a dynamic torsion, we propose an approximation scheme to obtain the contribution of the torsion in a perturbative manner and express the equations of the theory in terms of the pseudo-Riemannian quantities. In Section 5, we briefly discuss Dirac spinor coupling to the Einstein-Gauss-Bonnet gravitational Lagrangian. In sect. 6,  we consider the maximally symmetric de-Sitter  spacetime, which is compatible with past cosmological constant dominated universe with exponential expansion during inflationary era and also present accelerated expansion epochs of the universe.  We calculate the field equations for both pseudo-Riemannian and non-Riemannian cases and we see that a nontrivial but simple a solution of the pseudo-Riemannian theory that persists for the model with the torsion as well.   The paper ends with some brief comments on the results.     

\section{Notation and Preliminaries}

\subsection{Notation and some geometric preliminary}

We consider a four dimensional manifold in which the space-time metric can be defined in an orthonormal frame as
\begin{equation}
g=\eta_{ab}\,\theta^a \otimes  \theta^b,
\end{equation}
where $\eta_{ab}=\mbox{diag}(-1,1,1,1)$ is the Minkowski metric and $\theta^a, a=0,1,2,3$ denote the orthonormal basis one-forms. Most of the time, tensor product operator $\otimes$ is omitted following the common abuse of notation. In general these forms satisfy  Cartan's first and second structure equations
\begin{eqnarray}
&&{\bf{D}} \theta^a=d\theta^a+{\bf{\Gamma}}^a_{\phantom{a}b}\wedge \theta^b\equiv \Theta^a,\\
&&{\bf \Omega}^a_{\phantom{a}b}=d{\bf \Gamma}^a_{\phantom{a}b}+{\bf \Gamma}^a_{\phantom{c}c}\wedge {\bf \Gamma}^{c}_{\phantom{c}b}, \label{nonRiemOmega}
\end{eqnarray}
respectively, where ${\bf D}={\bf D}({\bf\Gamma})$ is the covariant exterior derivative operator of a connection with torsion.
 $d$ is the exterior derivative operator, ${\bf \Gamma}^a_{\phantom{a}b}$ denote connection one forms. 
We shall assume that the connection 1-form is metric compatible: 
$\mathbf{D}\eta_{ab}=\eta_{ac}\mathbf{\Gamma}^{c}_{\phantom{a}b}+\eta_{bc}\mathbf{\Gamma}^{c}_{\phantom{a}a}=0$.
$\wedge$ denotes the antisymmetric tensor product, ${\bf \Omega}^a_{\phantom{a}b}$ denotes  the curvature 2-forms in a Riemann-Cartan geometry.
In place of the torsion 2-forms $\Theta^a$, one can equivalently use the antisymmetric tensor-valued contorsion 1-forms $ K_{ab}=-K_{ba}\equiv K_{abc}\theta^c$, which can be defined by
\begin{equation}
\Theta^a=\frac{1}{2} \Theta^a_{\phantom{a}bc}\,\theta^{b}\wedge \theta^c= K^a_{\phantom{b}b}\wedge\theta^b.
\end{equation}
For a vanishing torsion, we have a metric compatible and torsion-free connection 1-forms, namely the Levi-Civita connection 1-forms $\omega^a_{\phantom{a}b}$ to satisfy the Cartan's structure equations
\begin{eqnarray}
&&D \theta^a=d\theta^a+\omega^a_{\phantom{a}b}\wedge \theta^b=0,
\\
&&\Omega^a_{\phantom{a}b}=d\omega^a_{\phantom{a}b}+\omega^a_{\phantom{c}c}\wedge \omega^{c}_{\phantom{c}b}=\frac{1}{2}R^a_{\phantom{a}bcd}\,\theta^c\wedge\theta^d,
\end{eqnarray}
where $R^a_{\phantom{a}bcd}$ are  the components of the Riemann tensor relative to an orthonormal coframe.
In order to distinguish pseudo-Riemannian and non-Riemannian quantities, we shall use the boldface letters for the geometrical quantities in the Riemannian-Cartan geometry. 
We also make use of the shorthand notation
\begin{equation}
\theta^{ab...c}=\theta^a\wedge\theta^b\wedge...\wedge \theta^c
\end{equation}
for the exterior products of the basis 1-forms for notational simplicity.

In four dimensions, the volume element with a particular orientation is given by the Hodge dual (equivalently the left dual) of unity as
\begin{equation}
*1=\theta^{0}\wedge\theta^{1}\wedge\theta^{2}\wedge\theta^3.
\end{equation}
The interior product with respect to the basis frame field $e_a$ will be denoted by  $i_a\equiv i_{e_a}$ and it satisfies the following computationally useful identities
\begin{equation}
i^k*\theta^{ab\ldots c}=*\theta^{ab\ldots ck}, \quad\qquad \theta^a\wedge* \theta^{b_1 b_2\ldots b_p}=(-1)^{p+1}*i^{a}\left(\theta^{b_1b_2\ldots b_p} \right).
\end{equation}
where the indices of the (co)frame fields are raised/lowered by the metric $\eta^{ab}$ and $\eta_{ab}$ as well.

One can define the Ricci 1-forms $R^a$, Ricci scalar $R$ and Einstein 1-forms $G^a$ as the following contractions of the curvature 2-forms:
\begin{eqnarray} \label{def_curv}
R^a=R^a_{\phantom{a}b}\theta^b=i_b\Omega^{ba},\quad  R=i_a R^a,\quad  G^{a}=G^a_{\phantom{a}b}\theta^b=R^a-\frac{1}{2}R \theta^a,
\end{eqnarray}
respectively. We have  also the convenient contraction identity, which follows simply from the coframe variational derivative of the Einstein-Hilbert Lagrangian,  
\begin{equation}\label{ein-def}
*G^a=-\frac{1}{2}\Omega_{bc}\wedge*\theta^{abc}.
\end{equation}
The eq. (\ref{ein-def}) expresses the Hodge dual of the Einstein 1-form $G^a$ as a contraction of the curvature 2-forms with the basis 3-forms. Moreover,  formulas (\ref{def_curv}) and (\ref{ein-def}) are valid  both in the pseudo-Riemannian and in Riemann-Cartan geometries, where the corresponding  non-Riemannian counterparts can be obtained by appropriately replacing non-Riemannian curvature two forms (\ref{nonRiemOmega}) in the above equations.

Both the left and right duals of the curvature 2-form will be used in the expression of the  metric field equations that will be derived in the following section. Note that the right dual pertains to four spacetime dimensions only. The left dual of the curvature 2-form is simply defined to be identical to the definition of  the Hodge dual as
\begin{equation}
*\Omega^{ab}
=
\frac{1}{2}R^{ab}_{\phantom{aa}cd}*\theta^{cd}
=
\frac{1}{4}R^{ab}_{\phantom{aa}cd}\epsilon^{cd}_{\phantom{aa}mn}\theta^{mn}
\end{equation}
whereas the right dual can be defined   by the relation
\begin{equation}\label{left-dual-def}
\Omega^{ab}*
\equiv
\frac{1}{2}\epsilon^{ab}_{\phantom{aa}cd}\Omega^{cd}
=
\frac{1}{4}\epsilon^{ab}_{\phantom{aa}mn} R^{mn}_{\phantom{aa}cd}\theta^{cd}
\end{equation}
in terms of the permutation symbol $\epsilon_{abcd}$.
In other words, the right dual is defined to act on the first pair of the indices  of the Riemann tensor.
These definitions are in accordance with the right/left duals defined, for example, in \cite{exact-solutions}
with respect to the indices of the Riemann tensor.

Likewise, by applying both duality definitions, the double-dual of the curvature 2-form, denoted for simplicity by $\tilde{\Omega}^{ab}$ is defined as
\begin{equation}
*\Omega^{ab}*
=
\frac{1}{8}\epsilon^{ab}_{\phantom{aa}mn} R^{mn}_{\phantom{aa}rs}\epsilon^{rs}_{\phantom{aa}cd}\theta^{cd}.
\end{equation}

The dual curvature tensors have some peculiar properties. For instance, the Einstein 3-form (\ref{ein-def}) can alternatively be expressed in terms of left dual curvature as
\begin{equation}
*G^a
=
-\Omega^{a}_{\phantom{a}b}*\wedge \theta^b,
\end{equation}
and consequently the Einstein 3-form corresponding to the curvature 2-form satisfying $\Omega^{a}_{\phantom{a}b}*=\mp\Omega^{a}_{\phantom{a}b}$ vanishes identically
\cite{straumann-gr-book}.

The field equations arising from a Gauss-Bonnet term, as we shall explicitly present in the next section, can be written in a very convenient form in terms of the double-dual curvature tensor which is particularly well-suited to the use of exterior algebra of forms. For further technical details and the interrelations involving the duals of the curvature 2-form in four dimensions, the reader is referred to \cite{benn-tucker,straumann-gr-book}.

\subsection{Dimensionally-continued Euler-Poincar\`{e} forms }

In order to introduce the Lagrangian density in a convenient  form, we now briefly mention the Gauss-Bonnet terms expressed also in terms of 
exterior algebra of forms in this section.

The famous Chinese  differential geometer S-S. Chern used dimensionally-continued Euler-Poincar\'{e} forms to generalize the
Gauss-Bonnet theorem to higher dimensions \cite{chern}.
The Euler characteristic $\chi(M)$ of a four dimensional, compact and oriented manifold $M$ can be expressed in terms of the dimensionally-continued Euler-Poincare form of second degree as
\begin{equation}
\chi(M)
=
\frac{1}{32\pi^2}\int_M \Omega_{ab}\wedge \Omega_{cd}*\theta^{abcd},
\end{equation}
which can be considered as a direct generalization of Gauss-Bonnet theorem in two dimensions relating the Gaussian curvature of a two dimensional closed manifold to one of its topological invariants. The generalized Euler characteristic  is defined
for an even dimensional manifold. However, the term ``Gauss-Bonnet" is more commonly used than, what one may call a more convenient  term,  the ``Euler-Poincare" in the literature.

According to the theorem by Lovelock \cite{lovelock1,lovelock2} bearing his name, the most general  gravitational Lagrangian density $n$-form leading to the field equations that are
of second order  in metric coefficients is of the form
\begin{equation}
L^{(n)}_{EP}=\sum^{[\frac{n}{2}]}_{k=0}\frac{\alpha_k}{2^k}\Omega_{a_1b_1}\wedge\cdots\wedge\Omega_{a_kb_k}\wedge *\theta^{a_1b_1\cdots a_kb_k},
\end{equation}
in $n$ dimensions where the square brackets in the upper limit of the summation denotes the integer value of the quantity inside the brackets. $\alpha_k$ is a coupling constant for the Euler-Poincare form of $k\mbox{-th}$ order. 
A general term in the sum can also be rewritten in an alternate form
\begin{equation}
\frac{1}{2^k}\Omega_{a_1b_1}\wedge\cdots\wedge\Omega_{a_kb_k}\wedge *\theta^{a_1b_1\cdots a_kb_k}
=
\frac{1}{2^{2k}}R_{a_1b_1}^{\phantom{a_1b_1}c_1d_1}\cdots R_{a_kb_k}^{\phantom{a_kb_k}c_kd_k}\delta^{a_1b_1\cdots a_kb_k}_{c_1d_1\cdots c_kd_k}*1,
\end{equation}
in terms of the product of $k$ number of Riemann tensor components and a generalized Kronecker symbol.

The first  few terms, corresponding to $k=0$, $k=1$, $k=2$ and $k=3$ in the sum can explicitly be rewritten in the form
\begin{eqnarray}
L^{(n)}_{EP}
&=
\lambda_0 *1+\frac{\alpha_1}{2}\Omega_{ab}\wedge*\theta^{ab}+\frac{\alpha_{2}}{4}\Omega_{ab}\wedge\Omega_{cd} \wedge *\theta^{abcd}
+\frac{\alpha_{3}}{8}\Omega_{ab}\wedge\Omega_{cd}\wedge \Omega_{ef}  \wedge *\theta^{abcdef} +\cdots.
\end{eqnarray}
The term for $k=0$  corresponds to a cosmological constant term whereas the term $k=1$ is the familiar Einstein-Hilbert term. The term with
$k=2$, the Gauss-Bonnet term, is a topological term in four dimensions and it can be expressed as an exact form as the total exterior derivative of a 3-form 
\cite{mielke} in the form
\begin{equation}
L^{(2)}_{EP}
=
dK,
\end{equation}
with the 3-form $K$ explicitly given by 
\begin{equation}
K=
\left(
\omega_{ab}\wedge \Omega^{cd}-\frac{1}{3}\omega_{ae}\wedge \omega^{e}_{\phantom{a}b}\wedge \omega^{cd}
\right)\epsilon^{ab}_{\phantom{ab}cd}.
\end{equation}
Consequently, the exact differential term $L^{(2)}_{EP}$ does not yield  field equations for the metric coefficients in four spacetime dimensions.
Note, on the other hand, that it yields an equations of motion for the metric coefficients in spacetime dimensions $n>4$.

A way to make such a term  yield some field equations in four spacetime dimensions is to couple
this term with a scalar field, or else consider a modified Lagrangian density of the form $f(\mathcal{G})$ where the scalar invariant $\mathcal{G}$ can be defined by the relation
\begin{equation}\label{GB-scalar-def}
\mathcal{G}*1
=
\frac{1}{4}\Omega_{ab}\wedge\Omega_{cd} \wedge *\theta^{abcd}.
\end{equation}

As  a side remark, also note that it is possible to write the $k=2$ term in an alternate form as
\begin{eqnarray}
\frac{1}{4}\Omega_{ab}\wedge\Omega_{cd}\wedge *\theta^{abcd}
&=
\frac{1}{2}\Omega_{ab}\wedge* \Omega^{ab}-R_a\wedge *R^a+\frac{1}{4}R^2*1
\\
&=
\frac{1}{4}(R_{abcd}R^{abcd}-4R_{ab}R^{ab}+R^2)*1,
\end{eqnarray}
in terms of the contraction of the Riemann tensor components.
This convenient relation implies that the  quadratic curvature Lagrangians in the expression on the right hand side are related by a surface term and
are not all independent of one another in four dimensions.
Consequently, in four spacetime dimensions the most general gravitational Lagrangian involving quadratic curvature terms can be written, for example, in the form
\begin{equation}
L_{qc}=a\, R^2*1+ b\, R_a \wedge* R^a
\end{equation}
involving only the scalar curvature and Ricci tensor.

The Gauss-Bonnet Lagrangians defined above have also been studied in the context of teleparallel gravity theories with
vanishing curvature tensor but having a non-vanishing torsion by introducing  a Weitzenb\"ock connection.
In arbitrary spacetime dimensions, the teleparallel equivalent of the Einstein-Hilbert Lagrangian is extended to case of the Gauss-Bonnet Lagrangians in a remarkable paper by Kofinas and Saridakis \cite{saridakis}.

\section{Gauss-Bonnet term in pseudo-Riemmannian case}

In this section, we shall consider a modified gravity theory with  Gauss-Bonnet terms  included in four spacetime dimensions as given below. Here, the Einstein-Hilbert Lagrangian 4-form is extended by (the Euler-Poincare form of second degree) the Gauss-Bonnet  term \cite{dereli-tucker-effective-string-models,wetterich,gasperini-giovannini}
\begin{equation}\label{lag-def}
L_{EGB}
=
\frac{1}{2\kappa^2}{\bf\Omega}_{ab}\wedge* \theta^{ab}-\frac{1}{2}d\phi\wedge *d\phi-V(\phi)*1+f(\phi)L_{GB}.
\end{equation}
with
\begin{equation}
{ L}_{GB}=\frac{\alpha}{4}{\bf\Omega}_{ab}\wedge {\bf \Omega}_{cd}*\theta^{abcd},
\end{equation}
where we have  $\alpha_2\equiv\alpha$ for simplicity in the GB Lagrangian and boldface or normal fonts will denote the terms  constructed from non-Riemannian or pseudo-Riemannian curvature 2-forms, respectively. Note that we assume at first  a general theory and impose  the metric compatible torsion free connection as a constraint. After taking the variation and subsequently by imposing the constraint, the field equations of the theory will be expressed solely in terms of pseudo-Riemannian quantities eventually. The model based on the Lagrangian 4-form (\ref{lag-def}) has two functions, the potential term for the scalar field,  $V(\phi)$ and a coupling function $f(\phi)$. For a dynamical scalar   field $\phi$,  the presence of a coupling function $f(\phi)$ will allow  $L_{GB}$ to contribute to the coframe field equations despite the fact that it is a total derivative in four dimensions.

In the constrained first order formalism we consider, recall that the independent connection is metric compatible,
 ${\bf\Gamma}_{ab}+{\bf\Gamma}_{ba}=0$. The vanishing torsion constraint, which is a dynamical constraint on the independent connection 1-form, can conveniently be implemented  by introducing a vector-valued Lagrange multiplier 2-form $\lambda^a$. Thus, the original Lagrangian (\ref{lag-def}) is extended to the form
\begin{equation}\label{extended-lag}
L'[\theta^a, {\bf \Gamma}^{a}_{\phantom{a}b}, \phi, \lambda^a]=L_{EGB}[\theta^a, {\bf \Gamma}^{a}_{\phantom{a}b},\phi ]
+
\lambda_a\wedge \left(d\theta^a+{\bf \Gamma}^{a}_{\phantom{a}b}\wedge \theta^b\right).
\end{equation}

In the calculation, we make essential use the following variational formulas:
\begin{equation}
\delta*\theta^{a\ldots b}=\delta\theta^c\wedge* \theta^{a\ldots b}_{\phantom{a\ldots b}c}, 
\quad \delta {\bf\Omega}^{ab}={\bf D}\delta{\bf \Gamma}^{ab},
\end{equation}
to obtain the total variational derivative, where $\delta$ denotes the variation of a quantity.
The total variational derivative of the extended Lagrangian (\ref{extended-lag}) with respect to its independent variables  can be written in the form
\begin{eqnarray}
&&\delta L'
=
\delta \theta_a\wedge
\left(
\frac{1}{2\kappa^2}\mathbf{\Omega}_{bc}\wedge *\theta^{abc}
+
*T^a[\phi]
-
V(\phi)*\theta^a+{\bf D}\lambda^a
\right)
\nonumber\\
&&\phantom{=aaa}+
\delta{\bf\Gamma}_{ab}
\wedge
\left[
-
\frac{1}{2}(\theta^a\wedge\lambda^b-\theta^b\wedge\lambda^a)
+
\frac{\alpha}{2\kappa^2}{\bf D}*\theta^{ab}
+
\frac{1}{2}{\bf D}
(f\mathbf{\Omega}_{cd}*\theta^{abcd})
\right]
\nonumber\\
&&\phantom{=aaa}
+
\delta\phi
\left(
d*d\phi-V'(\phi)*1+f'{ L}_{GB}
\right)
+
\delta\lambda_a\wedge \Theta^a,  \label{total-variational-der}
\end{eqnarray}
where  prime denotes the differentiation with respect to the scalar field and the energy-momentum form $*T^a[\phi]$ for the scalar field
can explicitly be written as
\begin{equation}
*T_a[\phi]
=
\frac{1}{2}(i_ad\phi*d\phi+d\phi \wedge i_a*d\phi).
\end{equation}
Considered as an equation for the Lagrange multiplier forms, the connection equation $\delta L'/\delta{\bf\Gamma}_{ab}=0$ can be
solved for $\lambda^a$ subject to the constraint $\Theta^a=0$ which follows from the field equations $\delta L'/\delta\lambda_{a}=0$. 
The constraint $\Theta^a=0$ then implies that the independent connection 1-form ${\bf \Gamma}_{ab}$ is the Levi-Civita connection.
Consequently, it is possible to assume that ${\bf D}$ corresponds to  the covariant exterior derivative of the Levi-Civita connection 1-forms, ${\bf D}=D$, and thus
the connection equations reduce to 
\begin{equation}\label{reduced-connection-eqns}
\frac{1}{2}(\theta^a\wedge\lambda^b-\theta^b\wedge\lambda^a)
=
\frac{\alpha}{2}f'd\phi\wedge \Omega_{cd}*\theta^{abcd}.
\end{equation}

Furthermore, the expression on the right-hand side of (\ref{reduced-connection-eqns}) can be rewritten in a more convenient form
by using, for example, the symmetry properties of the indices of the Riemann tensor $R_{abcd}=R_{cdab}$. Explicitly, one can readily derive the identity
\begin{equation}
\frac{1}{2}\Omega_{cd}*\theta^{abcd}
=
*\left[
\Omega^{ab}
-
(\theta^a\wedge R^b-\theta^b\wedge R^a)
+
\frac{1}{2}R\theta^{ab}
\right].
\end{equation}
On the other hand, by the definition of the left dual of the curvature 2-form (\ref{left-dual-def}), one obtains the identity
\begin{equation}
\Omega^{ab}*
=
*[
\Omega^{ab}
-
(\theta^a\wedge R^b-\theta^b\wedge R^a)
+
\frac{1}{2}R\theta^{ab}].
\end{equation}
In other words, the curvature expression on the right hand side in (\ref{reduced-connection-eqns}) is the right dual of the
double-dual curvature denoted by $\tilde{\Omega}^{ab}$ 
\begin{equation}
\tilde{\Omega}^{ab}
=
-\Omega^{ab}
+
(\theta^a\wedge R^b-\theta^b\wedge R^a)
-
\frac{1}{2}R\theta^{ab}. \label{doubledualcurvature}
\end{equation}

Now, returning to the independent connection equations (\ref{reduced-connection-eqns}), one can obtain a unique solution to the Lagrange multiplier 2-form 
expressed in terms of the double-dual curvature 2-forms.
Explicitly, calculating  two successive contractions of the connection equation (\ref{reduced-connection-eqns}) by taking the zero torsion constraint into account, one finds that the expression for Lagrange multiplier 2-form is of the form
\begin{equation}\label{lag-mult-explicit}
\lambda^b
=
-2\alpha\,f'(\phi) (i_{a}d\phi) *\tilde{\Omega}^{ab}.
\end{equation}

The expression (\ref{lag-mult-explicit}) is now to be used to eliminate the Lagrange multiplier term from the coframe variational derivative.
By using the second Bianchi identity in the form $D\Omega^{ab}=0$ in the pseudo-Riemannian case and inserting (\ref{lag-mult-explicit}) into the coframe  equations, the equations $\delta L'/\delta \theta_a\equiv *E^a $ can explicitly be rewritten in terms of the pseudo-Riemannian quantities.  Eventually, they take the compact form
\begin{equation}\label{reduced-coframe-eqn}
*E^a
=
-
\frac{1}{\kappa^2}*G^a
+
*T^a[\phi]
-
V(\phi)*\theta^a
+
2\alpha\,[f'' d\phi (i_{b}d\phi)  +f' D(i_{b}d\phi) ]\wedge *\tilde{\Omega}^{ab}
=0.
\end{equation}

In comparison with the previous work of Gasperini and Giovannini \cite{gasperini-giovannini}, we note that up to this point we have not resorted to an approximation 
method to derive the metric (equivalently, the coframe) field equations and that the Lagrange multiplier forms can successfully be eliminated from the metric equations by finding a unique solution to the connection equations in terms of the double-dual curvature 2-form. 
We also note that the term of the form $\Omega_{bc}\wedge \Omega_{de}\wedge *\theta^{abcde}$ is absent in four spacetime dimensions since this term survives only for the spacetime dimensions $n>4$.

The trace of the coframe equations can be calculated by taking the wedge product of (\ref{reduced-coframe-eqn}) from the left with the basis coframe 1-form $\theta_a$.
One explicitly finds
\begin{equation}
\frac{1}{\kappa^2}R*1
+
\frac{1}{2}d\phi\wedge *d\phi
-
4V(\phi)*1
-
d(\theta_a\wedge \lambda^a)=0.
\end{equation}
The last term can be simplified by noting that by the definition of the double-dual curvature (\ref{doubledualcurvature}), one has $i_a \tilde{\Omega}^{ab}=G^b$ where $G^b$ is the Einstein 1-form,
and therefore the trace can be written as
\begin{equation}
\frac{1}{\kappa^2}R*1
-
\frac{1}{2}d\phi\wedge *d\phi
-
4V(\phi)*1
+
2 \alpha 
[f'' d\phi (i_{a}d\phi)  +f' D(i_{a}d\phi) ]\wedge * G^a
=
0.
\end{equation}

Finally, the field equation for the scalar field obtained from the expression for the total the variational derivative (\ref{total-variational-der}) is
of the form
\begin{equation}\label{scalar-fe}
d*d\phi-V'(\phi)*1+f'L_{GB}=0.
\end{equation}
 Thus, without the kinetic term in the total Lagrangian 4-form (\ref{lag-def}), the field equation for the scalar field becomes more of an equation constraining the value of the Gauss-Bonnet terms and $\phi$ evidently is not a dynamic field in this case.

\subsection{A constrained $f(\phi)$ case \label{const-fphi}}

We note that the coframe field equations (\ref{reduced-coframe-eqn}) can be rewritten in terms of left dual curvature 2-form as 
\begin{equation}\label{reduced-coframe-eqn2}
*E^a
=
-
\Omega^{a}_{\phantom{a}b}*\wedge \left[\frac{1}{\kappa^2}\theta^b +\alpha D (i^bdf')\right]
+
*T^a[\phi]
-
V(\phi)*\theta^a
=0.
\end{equation}

With the coframe equations written in the form (\ref{reduced-coframe-eqn2}), it is easy to construct  an additional constraint  on the function $f$ 
that leads to a possible simplification in the resulting field equations. Explicitly, if one assumes that the term involving the  second derivatives of the function $f$ satisfies the constraints
\begin{equation}
\label{gurses-assumption}
2\alpha D (i^bdf')=\Lambda_1(\phi) \theta^b,
\end{equation}
for an arbitrary scalar field $\Lambda_1$, the coframe equations (\ref{reduced-coframe-eqn}) reduce to the following equation
\begin{equation}\label{reduced-gurses-scalar}
*E^a
=
\left[\Lambda_1(\phi)-\frac{1}{\kappa^2}\right]*G^a
+
*T^a[\phi]
-
V(\phi)*\theta^a
=0,
\end{equation}
The additional assumption (\ref{gurses-assumption}) has recently  been introduced by by G\"urses in \cite{gurses} where he also made use of  such simplifiying  assumption to find  exact solutions to the GB extended gravitational model (\ref{lag-def}) in four spacetime dimensions by adopting a conformally flat metric ansatz. A further simplification  occurs if one sets $\Lambda_1$ to a constant as 
\begin{equation}
\label{f-constraint}
\Lambda_1=\frac{1}{\kappa^2}(\lambda^2-1),
\end{equation}
for some dimensionless constant $\lambda$. Then the coframe equations (\ref{reduced-coframe-eqn}) reduce to the following equation for the metric for an Einstein-scalar system
\begin{equation}\label{reduced-einstein-scalar}
*E^a
=
-\frac{1}{\kappa'^2}*G^a
+
*T^a[\phi]
-
V(\phi)*\theta^a
=0,
\end{equation}
with a shifted gravitational coupling constant $\kappa'\equiv{\kappa}/{\lambda}$. 
Thus, the above considerations show that the GB extended model reduces to the Einstein-scalar model under some relaxed assumptions.  In particular, eq. (\ref{reduced-einstein-scalar}) is obtained imposing a constraint on the metric only indirectly through the 
connection involved in eqns. (\ref{gurses-assumption}) and (\ref{f-constraint}). It is interesting to investigate whether it is  possible to construct  solutions to the eqns. (\ref{reduced-einstein-scalar}) by using the known solutions of the Einstein-scalar gravity theory either with  a potential term or without potential by discarding potential term.  However, the scalar field equation (\ref{scalar-fe}) prevents this, due to the last term in the scalar field equation unless the last term vanishes. This is possible if either (i) $f'=0$ or (ii) $L_{GB}=0$. In case (i), we have a constant scalar and therefore the GB term does not contribute to the field equations, and therefore theory reduces to GR. In case (ii), we have a non-trivial scalar field but the GB invariant vanishes identically and the theory becomes Einstein-scalar theory with a potential. Note that under assumptions specified above, the modified GB model we consider  reduces to some other theories, and hence shares same  solutions, but in a  perturbative treatment, they may have contrasting properties.

\subsection{Modified GB extended model with a constrained $f(\mathcal{G})$ }

The additional constraint (\ref{gurses-assumption}) for the function $f(\phi)$ can also be extended to the modified Gauss-Bonnet Lagrangian  involving 
an algebraic function of the GB scalar of the form $f(\mathcal{G})$. The modified GB model is of considerable interest because it is 
a good candidate to emulate the dark energy in a cosmological context.

For simplicity, to discuss this point of view in connection with the constraint of type (\ref{gurses-assumption}) qualitatively, 
let us briefly consider the modified GB Lagrangian \cite{nojiri-odintsov} of the form
\begin{equation}
L_{mGB}
=
\frac{1}{2\kappa^2}R*1+f(\mathcal{G})*1,
\end{equation}
which has been studied by Nojiri and Odintsov to introduce a model featuring what one may call  ``gravitational dark energy". 

In the notation of the current work the metric field equations $*E^a=0$ that follow from the coframe variational derivative 
$\delta L_{mGB}/\delta \theta_a\equiv *E^a$
can be written in terms of  the 3-form $*E^a$ of the explicit form \cite{nojiri-odintsov,baykalunified}
\begin{equation}\label{mGB-equations}
*E^a
=
-\frac{1}{\kappa^2}*G^a+2D(i_bdf')\wedge *\tilde{\Omega}^{ba}-(f'\mathcal{G}-f)*\theta^a,
\end{equation}
where prime denotes a derivative with respect to the scalar $\mathcal{G}$ and the terms involving the second order derivatives of the 
function $f(\mathcal{G})$ arise, as in the previous cases from the Lagrange multiplier term imposing the vanishing torsion constraint. 
$f(\mathcal{G})$ in this case acts like a scalar  field lacking a corresponding kinetic term.  

Similarly to  the previous case, now assume that  the second order derivative of $f(\mathcal{G})$ satisfies
\begin{equation}
2 D(i^bdf')
=
\frac{1}{\kappa^2}h(\mathcal{G})\theta^b
\end{equation}
for some function $h(\mathcal{G})$. Consequently, the metric field equations (\ref{mGB-equations}) then reduce to
\begin{equation}\label{reduced-mGB-equations}
\frac{1}{\kappa^2}\left[1+h(\mathcal{G})\right]*G^a+V(\mathcal{G})*\theta^a=0
\end{equation}
where $V(\mathcal{G})=f'\mathcal{G}-f$ is the Legendre transform of the function $f(\mathcal{G})$. The reduced field equations 
(\ref{reduced-mGB-equations}) in this case correspond to a particular scalar-tensor model involving two independent functions 
 $V(\mathcal{G})$ and $h(\mathcal{G})$. In the subcase where $h(\mathcal{G})$ is a constant as in the preceding subsection, 
the modified GB model then simply reduces to  GR with a variable cosmological constant-like term.

\section{The case with non-vanishing Torsion}

The field equations that follow from (\ref{lag-def}) can be considered as a kind of  generalization of the Brans-Dicke gravity to the Gauss-Bonnet gravity
involving scalar field coupled nonminimally to the GB  scalar.
As in the original Brans-Dicke field equations, it is possible to study the field equations  assuming, for example, that the connection is metric compatible but not torsion-free. This case can be recovered from the independent variational derivative expression (\ref{total-variational-der}) by dropping the vanishing torsion constraint by setting $\lambda^a=0$.
Recall that we use the symbol ${\bf \Gamma}_{ab}$ for the independent non-Riemannian connection with torsion  relative to an orthonormal coframe. 
In accordance with this convention, the exterior covariant derivative with respect to the non-Riemannian connection, and the curvature 2-form corresponding to this connection will be denoted by the boldface letters ${\bf D}$ and ${\bf \Omega}^{ab}$ respectively, whereas for pseudo-Riemannian ones we use normal letters. By making these arrangements, the field equations  of the Lagrangian (\ref{lag-def}) for the non-Riemannian case  can be expressed  by considering  (\ref{total-variational-der}) as 
\begin{eqnarray}
 &&\frac{1}{\kappa^2}*{\bf G}^a(\mathbf{\Gamma})=  *T^a[\phi] - V(\phi)*\theta^a,\label{field-eq-nr1}
 \\
 &&{\bf D}*\theta^{ab} = -\kappa^2 \alpha\, {\bf D} \left[f\,{\bf \Omega}_{cd}*\theta^{abcd} \right],\label{torsion-eq}
 \\
&&d*d\phi-V'(\phi)*1+\frac{\alpha}{4}f'\,{\bf \Omega}_{ab}\wedge {\bf \Omega}_{cd}*\theta^{abcd}=0.\label{scalar-fe-tor}
\end{eqnarray}  
These are the field equations of the non-Riemannian  with torsion  expressed in terms of the non-Riemannian quantities for the metric field, the independent 
connection 1-forms with torsion and the scalar field, respectively.

 In the similar non-Riemannian Brans-Dicke case, the field equations can also be expressed in terms  of pseudo-Riemannian quantities, thanks to the torsion being algebraic. Now we follow same route and try to express the torsion forms in terms of the Riemannian quantities. To this end, we need to  
put  the equation (\ref{torsion-eq}) in a more convenient form.  By using the expression for the covariant exterior derivatives of the Hodge duals of the basis $p$-forms expressed in the form
\begin{equation}
{\bf D}
*\theta^{ab}
=
 \Theta_c\wedge *\theta^{abc},
\end{equation}
it is possible to find the equation satisfied by the torsion 2-form as 
\begin{equation}
 \Theta^c\wedge *\theta^{ab}_{\phantom{ab}c}
=
-\kappa^2\alpha \, df \wedge {\bf \Omega}_{cd}*\theta^{abcd}.
\end{equation}
By using the contraction identity $\theta_d\wedge*\theta^{abcd}=*\theta^{abc}$ in four dimensions, the above equation can be put into the form 
\begin{equation}\label{Torsion-eq0}
\theta_a\wedge  \Theta_b-\theta_b\wedge \Theta_a  =2\kappa^2\alpha\, df(\phi)\wedge {\bf\Omega}_{ab}. 
\end{equation}
By contracting this equation with first $i_b$ and then $i_a$  and using the identity $\theta^a \wedge i_a Q=p\, Q$  for the p-form $Q$, we obtain the following useful equation
\begin{equation}
 \Theta^a=2\kappa^2\alpha \left[ i_{b} \left(df\wedge {\bf\Omega}^{ba} \right)
+
\frac{1}{4} \theta^a\wedge  i_c i_{d}\left(df\wedge {\bf \Omega}^{cd} \right) \right]. \label{Torsion-eq}
\end{equation}
From the equation above, we can also obtain an expression for the contorsion one forms. By making use of the equation
\begin{equation}
 K_{ab}=-\frac{1}{2}i_a i_b \left( \Theta^c\wedge \theta_c \right)+i_a \Theta_b-i_b  \Theta_a,
\end{equation}
and  the torsion expression  (\ref{Torsion-eq}), one finds
\begin{eqnarray}
 K_{ab}&=\kappa^2\alpha\bigg\{   i_a i_b  i_{c} \left(df\wedge{\bf \Omega}^{cd}\wedge\theta_d \right)   
+  2  i_a i_{c} \left(df\wedge  {\bf \Omega}_{\phantom{b}b}^{c} \right)-  2 i_b i_{c} \left(df\wedge {\bf \Omega}_{\phantom{a}a}^{c}\right)\nonumber
\\ 
&\phantom{=\kappa^2\alpha\quad}+\frac{1}{2} (\theta_a\wedge  i_b -\theta_b\wedge i_a)\left[i_c i_{d}\left(df\wedge {\bf \Omega}^{cd} \right)\right] \bigg\}.
\label{kontorsion-eq}
\end{eqnarray}
 
The next step is to express the non-Riemannian curvature two-form ${\bf \Omega}^{ab}$  in terms of Riemannian one $\Omega^{ab}$ and contorsion one form. 
It is a convenient and practical custom to  decompose the non-Riemannian connection $\bf{\Gamma}$ into a pseudo-Riemannian part and a part coming from
torsion as
\begin{equation}\label{connection-decomp}
{\bf \Gamma}^{a}_{\phantom{a}b}
=
\omega^{a}_{\phantom{a}b}+ K^{a}_{\phantom{a}b}
\end{equation}
in a Riemann-Cartan geometry.
The decomposition (\ref{connection-decomp}) allows one to decompose any geometrical expression involving curvature into Riemannian parts and the parts
originating from the non-vanishing torsion.
For example, the curvature 2-form  can be decomposed as
\begin{equation}\label{curvature-decomp}
{\bf \Omega}^{a}_{\phantom{a}b}
=
\Omega^{a}_{\phantom{a}b}+D  K^{a}_{\phantom{a}b}+ K^{a}_{\phantom{a}c}\wedge  K^{c}_{\phantom{a}b}.
\end{equation}

However,  one can observe that the connection equations yield a set of dynamical equations for the torsion 2-form (\ref{Torsion-eq}), or equivalently the contorsion 1-forms (\ref{kontorsion-eq}).  Unfortunately, the resulting equations are not a set of algebraic equations for the contorsion 1-forms owing to 
the presence of the terms
\begin{equation}
D K^{a}_{\phantom{a}b}
=
dK^{a}_{\phantom{a}b}
+
\omega^{a}_{\phantom{a}c}\wedge K^{c}_{\phantom{a}b}
-
\omega^{c}_{\phantom{a}b}\wedge  K^{a}_{\phantom{a}c}.
\end{equation}
 The resulting field equations thus allow a propagating torsion as they stand in general.
Moreover, note also that $K_{ab}=0$ is not a solution to the connection equations and consequently it is not consistent to set the contorsion 1-forms
to zero in the coframe equations as well. 

As a consequence of the above results for the pseudo-Riemannian cases one can conclude that the nonminimally-coupled scalar field in the Gauss-Bonnet case poses some additional technical difficulties in the case with non-vanishing torsion due to the presence of the curvature terms 
arising in the variational derivatives of the independent connection 1-forms. Thus, one can resort to an approximation method \cite{gasperini-giovannini} to study
the field equation that follows from (\ref{lag-def}) in the context of the Riemann-Cartan case with non-vanishing torsion.  We start the approximation scheme with that of the first order in the GB coupling constant $\alpha$ in the next section. The technical details of the more involved calculations leading to metric equations that are the to second order in the coupling constant  are relegated to the appendix.

\subsection{The field equations in the $O(\alpha)$ approximation}

In order to  see the effect of the torsion we can use an approximation  by assuming that it has an expansion into  a series in the powers of the coupling constant $\alpha$. The first order approximation is then obtained by keeping the linear terms in the coupling constant $\alpha$ in equations (\ref{Torsion-eq}). This can be done, in the linear order in $\alpha$ and by assuming that the approximation 
 $
{\bf \Omega}^a_{\phantom{a}b} \simeq \Omega^a_{\phantom{a}b}
$ 
 can legitimately be inserted into  the field equations. Applying this approximation scheme, the torsion equation (\ref{Torsion-eq})  satisfies $\Theta^a=\Theta^{a(1)} +  $ higher order terms in $\alpha$, where to first order in $\alpha$ one has
   \begin{eqnarray}
\Theta^{a(1)}
&=&2\kappa^2\alpha \left[ i_{b} \left(df\wedge \Omega^{ba} \right)+\frac{1}{4} \theta^a\wedge  i_c i_{d}\left(df\wedge \Omega^{cd} \right) \right],\label{Torsion-eq1}
\end{eqnarray}
and this expression consequently  yields the interesting result
\begin{eqnarray}
\Theta^{a(1)}&=&2\kappa^2\alpha \, (i_{b}df)\left(\Omega^{ba}- \theta^b \wedge R^{a}+ \frac{1}{2} \,\theta^a\wedge R^b-\frac{1}{4}R\theta^a \wedge \theta^b\right). 
\end{eqnarray}
  
Likewise, the contorsion forms can be found by making use of the expression (\ref{kontorsion-eq}) and to first order in $\alpha$ they become 
\begin{eqnarray}
K_{ab}^{(1)}=  2\kappa^2\alpha\bigg\{    
   i_a i_{c} \left(df\wedge  \Omega_{\phantom{b}b}^{c} \right)-   i_b i_{c} \left(df\wedge \Omega_{\phantom{a}a}^{c}\right) 
+\frac{1}{4} (\theta_a\wedge  i_b -\theta_b\wedge i_a)\left[i_c i_{d}\left(df\wedge \Omega^{cd} \right)\right]
\bigg\},
\label{kontorsion-eq-1}
\end{eqnarray}
This expression can be simplified further to a convenient form as
\begin{eqnarray}\label{K1}
K_{ab}^{(1)}&=& 2\kappa^2\alpha \,  (i_{c}df)\left[  i^c \Omega_{ab} + R_{a}\, \delta_b^c   -  R_{b} \, \delta_a^c  +\frac{1}{2} \theta_a \,  G^c_b -\frac{1}{2} \, \theta_b \, G^c_a  \right].
\end{eqnarray}
The expression for the contorsion obtained above  is now to be inserted into the non-Riemannian field equation (\ref{field-eq-nr1}). Since the right hand side is free of torsion terms, we only have to calculate the Einstein three-form term $*{\bf G}^a$  on the left hand side.  In the approximation to the first order, we have
\begin{eqnarray}
*{\bf G}^a&=&
-\frac{1}{2}\left[ \Omega_{bc}+D K_{bc}+K_a^{\phantom{a}d}\wedge K_{dc}\right] \wedge *\theta^{abc} 
\nonumber\\
&\simeq& *G^a+D K_{bc}^{(1)}\wedge *\theta^{abc} .\label{G_firstorder}
\end{eqnarray}
Now, we calculate the second term above. Note that the terms of the form $K\wedge K$  above have been discarded, since they are second order in $\alpha$. In order to simplify this term,  we first calculate the following 
\begin{eqnarray} 
 K_{bc}^{(1)} \wedge * \theta^{abc}   &=& 4\kappa^2\alpha \,  (i_{d}df)  *\left[ -\Omega^{ad} + \theta^a \wedge R^d  
-   \theta^{d} \wedge R^a   -\frac{1}{2} R\, \theta^{ad}
    \right]
    \nonumber\\
        &=& 4\kappa^2\alpha \,  (i_{d} df)  * \tilde{\Omega}^{ad}, 
\end{eqnarray}  
where $\tilde{\Omega}^{ab}$ is the double-dual curvature form defined in (\ref{doubledualcurvature}). Hence,  by considering  (\ref{G_firstorder}), in the first order approximation in $\alpha$,  the coframe equations (\ref{field-eq-nr1}) become  
\begin{equation}\label{riemann-cartan-coframe-eqn}
\frac{1}{\kappa^2}* G^{a}=*T^a[\phi]-V(\phi)*\theta^{a}+ 2\alpha\,  \left[f''  d\phi\, i_b(d\phi)  +f' D(i_b d\phi) \right]\wedge  * \tilde \Omega^{ab}.  \label{fe1}
\end{equation} 

If one compares the coframe equations (\ref{riemann-cartan-coframe-eqn}) above to those of the pseudo-Riemannian case given in (\ref{reduced-coframe-eqn}), one concludes  that in the linear approximation the field equations turn out to be the same. Consequently, the effect of a nonvanishing torsion generated  cannot affect the field equations to be distinguished from Riemannian one at this order and that such effects emerge at higher order terms. Note that this fact is in accordance with EC theory having a spinning matter source, since when the quadratic terms due to spin are neglected,  EC and GR governed by identical metric field equations \cite{trautman}. 

One can show that the field equations for the scalar field (\ref{scalar-fe-tor}) for in the first order in $\alpha$ is of the form
\begin{equation}
d*d\phi-V'(\phi)*1+\frac{\alpha}{4}f'\Omega_{ab}\wedge \Omega_{cd}*\theta^{abcd}=0,
\end{equation} 
and  this  is also the same  as the corresponding expression in the Riemannian case as well. These results are compatible with the work \cite{gasperini-giovannini} in which they were also made a first order variation of a similar Lagrangian without a potential term in $n$-dimensions with torsion   but they keep their calculation at the linear order as in this subsection and they found that  the field equations agree with the pseudo-Riemannian case given in  \cite{wetterich}.  

The interesting result to the order $O(\alpha)$ also  calls for an expression to order  $O(\alpha^2)$  to have a tangible effect of a nonvanishing torsion to emerge  through  the  approximation scheme above. 

\subsection{Field equations to order $O(\alpha^2)$}

We extend our approximation to include the correction terms  up to the factor $\alpha^2$ in the field equations as follows. To do this we have to recalculate the  equations  (\ref{Torsion-eq}) or (\ref{kontorsion-eq}) but  the  ${\mathbf\Omega}_{ab}$ term in those equations is now to be replaced by
\begin{equation} 
{\bf \Omega}^{ab} \simeq  \Omega^{ab}+D K^{ab(1)},\label{omega-1}
\end{equation}
where $K^{ab(1)}$ is the contorsion one forms first order in $\alpha$ and given by  (\ref{kontorsion-eq-1}). If we replace the above equation into the equation (\ref{kontorsion-eq}), the second order correction to the contorsion 1-forms becomes 
\begin{eqnarray}
K_{ab}^{(2)}&=\kappa^2\alpha\bigg\{   i_a i_b  i_{c} \left(df\wedge D K^{cd(1)}\wedge\theta_d \right)   
+  2  i_a i_{c} \left(df\wedge D K_{\phantom{b}b}^{c(1)} \right)-  2 i_b i_{c} \left(df\wedge D K_{\phantom{a}a}^{c(1)}\right)\nonumber
\\ 
&\phantom{=\kappa^2\alpha\quad}+\frac{1}{2} (\theta_a\wedge  i_b -\theta_b\wedge i_a)\left[i_c i_{d}\left(df\wedge D K^{cd(1)} \right)\right] \bigg\}.\label{kontorsion-eq-2}
\end{eqnarray}

  After calculating  these $K^{ab(2)}$ terms, we have to feed them into the field equations (\ref{field-eq-nr1}) and (\ref{scalar-fe-tor}). Hence we have to calculate the Einstein one-forms $*{\bf G}^a$. Note that the $K\wedge K$
 terms that we have discarded in the previous subsection will  contribute to the field equations at this order. Thus, if we label the correction terms  in the order $\alpha^2$ which must be added  to the right hand side of the field equation (\ref{fe1}) expressed conveniently in terms of a vector-valued perturbation 1-forms $H^{a(2)}$, then  the metric field equations (\ref{field-eq-nr1}) become
 \begin{equation}
 \frac{1}{\kappa^2}* G^{a}=*T^a[\phi]-V(\phi)*\theta^{a}+ 2\alpha\,   \left[f''  d\phi\, i_b(d\phi)  +f' D(i_b d\phi) \right]\wedge  * \tilde \Omega^{ab}
 +*H^{a(2)},  \label{fe2}
 \end{equation}
 where we have   
\begin{equation} 
*H^{a(2)}=\frac{1}{2\kappa^2} \left[ K_{bd}^{(1)}\wedge K^{d(1)}_{\phantom{a}c}\wedge *\theta^{abc}+D K_{bc}^{(2)}\wedge *\theta^{abc}  \right].\label{H2}
\end{equation}
The expressions for these terms are long and not particularly illuminating, the explicit  expressions for them are given in the appendix for convenience of the reader.

In this order, scalar field eq. (\ref{scalar-fe}) also gets a modification. As a consequence of the  expression (\ref{omega-1}), one can show that eq. (\ref{scalar-fe}) takes the form
\begin{equation}\label{scalara2}
d*d\phi-V'(\phi)*1+\frac{\alpha}{4}f'  \Omega_{ab}\wedge  \left(\Omega_{cd}+ 2\, D K_{cd}^{(1)}\right)*\theta^{abcd}=0
\end{equation} 
where $DK_{ab}^{(1)}$ term is given in (\ref{DK1}) below in Appendix.

\subsection{$O(\alpha^n)$-order approximation}

Iterating  the above approximation scheme by a desired number of times, and also by using an expansion for the non-Riemannian curvature 2-forms of the form
\begin{equation}\label{omega-n}
{\bf \Omega}^{ab}=\Omega^{ab}+\sum_{k=1}^n\left[ DK^{ab(k)}+\left(K^{a}_{\phantom{a}c}\wedge K^{cb}\right)^{(k)}\right], 
\end{equation}
 we can, in principle, calculate the metric field eq. (\ref{field-eq-nr1}) to $O(\alpha^n)$ as given by
\begin{equation}
 \frac{1}{\kappa^2}* G^{a}=*T^a[\phi]-V(\phi)*\theta^{a}+ 2\alpha\,  \left[f''  d\phi\, i_b(d\phi)  +f' D(i_b d\phi) \right]\wedge  * \tilde \Omega^{ab}
 +\sum_{k=2}^n*H^{a(k)},  \label{fen}
 \end{equation}
with
\begin{eqnarray} 
*H^{a(k)}=\frac{1}{2\kappa^2} \left[\left( K_{bd}\wedge K^{d}_{\phantom{a}c}\right)^{(k)}\wedge *\theta^{abc}+D K_{bc}^{(k)}\wedge *\theta^{abc}  \right],
\end{eqnarray}
where the first term above can be constructed from the terms of the form  $K\wedge K$ in the order $\alpha^k$.
Note that the $K_{bc}^{(k)}$ terms must be constructed recursively from the equation
\begin{eqnarray}
&&K_{ab}^{(k)}=\kappa^2\alpha\bigg\{   i_a i_b  i_{c} \left(df\wedge DK^{cd(k-1)}\wedge\theta_d \right)   
+  2  i_a i_{c} \left(df\wedge D K_{\phantom{b}b}^{c(k-1)} \right)\nonumber
\\ 
&&\phantom{=\kappa^2\alpha aaaaaa }-  2 i_b i_{c} \left(df\wedge D K_{\phantom{a}a}^{c(k-1)}\right)+\frac{1}{2} (\theta_a\wedge  i_b -\theta_b\wedge i_a)\left[i_c i_{d}\left(df\wedge D K^{cd(k-1)} \right)\right] \bigg\}.\label{kontorsion-eq-n-2}
\end{eqnarray}
with the help of $K_{ab}^{(k-1)}$ terms. Moreover, the scalar field eq. (\ref{scalar-fe-tor}) must also be handled to include such correction terms in the Gauss-Bonnet Lagrangian by using the eq. (\ref{omega-n}).

\section{Minimally coupled Dirac field}

Besides the torsion generated by the nonminimal couplings of matter field arising, for example, from the scalar fields coupling to a curvature tensor components,
the spinor fields are also known to generate torsion in a perhaps more natural way by the well-known minimal coupling prescription in the context of  Riemann-Cartan geometry. We briefly discuss spinor couplings starting from the Lagrangian governing the field equations for the Dirac spinor $\psi$ after 
introducing the basic notation required to write out the Lagrangian and the corresponding field equations  in a form convenient to the notation used above.

The exterior algebra of tensor-valued forms, which are adopted in the current work, can naturally  be extended to Clifford algebra-valued forms 
\cite{clifforms-mielke,topological-action-mielke} to discuss the spinor  couplings to (modified) gravity. We shall use the notation introduced in ref. \cite{dereli-tucker-spinor-bd} which discusses the spinor coupling to the geometry in the context of the Brans-Dicke theory with torsion induced by the BD scalar field.

For the Dirac spinor $\psi$ (spinor-valued 0-form), the Lagrangian density can be written in the form
\begin{equation}\label{dirac-lag}
L_D[\psi, \bar{\psi}, \theta^a, \mathbf{\Gamma}_{ab}]
=
\frac{1}{2}
\left(
\bar{\psi}*\theta^a \wedge \gamma_a \mathbf{D}\psi+\overline{\mathbf{D}\psi}\wedge *\theta^a \gamma_a \psi
\right)
-
m\bar{\psi}*\psi,
\end{equation}
where $m$ stands for the mass of the Dirac particle, whereas a bar over a field refers to the  conjugate 4-component Dirac spinor. The prescription of the minimal coupling to the gravity  simply requires  $d\psi$ and $d\bar{\psi}$ terms on the flat Minkowski background are to be replaced by the 
$SL(2,\mathbb{C})$-covariant exterior derivatives $\mathbf{D}\psi$ and $\overline{\mathbf{D}\psi}$ respectively for the curved background spacetime. In terms of the Riemann-Cartan connection 1-forms 
$\mathbf{\Gamma}_{ab}$, the action of the covariant exterior derivative on a Dirac spinor $\psi$  has the explicit form  
\be
\mathbf{D}\psi
=
d\psi
+\frac{1}{2}\mathbf{\Gamma}^{ab} \sigma_{ab}\psi,\qquad 
\overline{\mathbf{D}\psi}
=
d\bar{\psi}
-
\frac{1}{2}\mathbf{\Gamma}^{ab} \bar{\psi}\sigma_{ab},
\ee
with the generators $\sigma_{ab}$ defined in terms of  the commutator of the $4\times 4$ gamma-matrices as
\be
\sigma_{ab}
\equiv
\frac{1}{4}[\gamma_a,\gamma_b]_-.
\ee
Thus, by definition $\mathbf{D}$ acting on a Dirac spinor $\psi$ yields a matrix-valued 1-form, whereas $\mathbf{D}^2\psi=\frac{1}{2}\mathbf{\Omega}^{ab}\sigma_{ab}\psi$.
The set of the gamma matrices $\{\gamma_a\}$ with the indices referring to an orthonormal coframe generates the Clifford algebra. The indices of the gamma-matrices are raised/lowered by the flat metric $\eta_{ab}$ and they have the defining anti-commutation relations expressed explicitly in of the form
\be
\left[\gamma_a,\gamma_b\right]_+
=
\gamma_a\gamma_b+\gamma_b\gamma_a=2\eta_{ab}I_{4\times 4}
\ee
where $I_{4\times 4}$ is the unit $4\times 4$ matrix. With the metric signature assumed to be mostly plus, we have $\gamma^\dagger=\gamma^0\gamma^\mu\gamma^0$
and $\bar{\psi}=i\psi^\dagger \gamma^0$. The Clifford algebra generators $\sigma_{ab}$ satisfy the same commutation relations as the generators of the Lie algebra of the Lorentz group $SO(1,3)\simeq SL(2,\mathbb{C})$.

One can show that the variational derivative of the $SL(2,\mathbb{C})$ gauge covariant Lagrangian (\ref{dirac-lag}) with respect to the independent field variables yields
\begin{eqnarray} 
\delta L_D
&&=
\delta\bar{\psi}
\left[
*\theta^a\wdg \gamma_a \mathbf{D}\psi
-
\frac{1}{2}(\mathbf{D}*\theta^a)\gamma_a\psi
-
m*\psi
\right]
+
\left[
\overline{\mathbf{D}\psi} \gamma_a\wdg *\theta^a
-
\frac{1}{2}\bar{\psi} \gamma_a \mathbf{D}*\theta^a
-
m*\bar{\psi}
\right]\delta\psi
\nonumber\\
&&\fant{=}+
\delta \theta^a
\wdg \left[
\frac{1}{2}*\theta^b_{\fant{a}a} \wedge \left(\bar{\psi}\gamma_b \mathbf{D}\psi-\overline{\mathbf{D}\psi}\gamma_b \psi\right)
-
m\bar{\psi}\psi*\theta_a
\right]
+\delta \mathbf{\Gamma}^{ab}\wdg 
\frac{1}{2}*\theta^c \bar{\psi}\left[\gamma_c,\sigma_{ab}\right]_+ \psi,
\end{eqnarray}
up to an omitted exact form. 

In the context of the Riemann-Cartan geometry,
we first note that the field equation for $\psi$ is to be modified by a torsion term involving $\mathbf{D}*\theta^a=\Theta^b\wdg *\theta^{a}_{\fant{a}b}$
coupled to the gravity as well. Unfortunately, in the case of a dynamic torsion obtained, for example, from the total Lagrangian of the form $L_{tot.}=L_{EGB}+L_D$ leads to a set of fairly complicated,  coupled field equations rendering an analytical treatment to find an exact solution difficult. 
The field equations that follow from Einstein-Hilbert Lagrangian supplemented by $L_D$ can be expressed in terms of pseudo-Riemannian quantities, since in this case the algebraic torsion leads to  quartic spinor-spinor  and spinor-gravitational interaction terms.  Similar considerations have also been introduced  in the Brans-Dicke theory \cite{dereli-tucker-torsion}. On the other hand, in the case of gravitational part involving $L_{EGB}$ leads to a dynamic torsion field.
Explicitly, by combining the variational derivatives of the Lagrangians calculated above, one can readily obtain  from the variation with respect to an independent connection $\delta L_{tot.}/\delta\mathbf{\Gamma}_{ab}=0$  an equation involving torsion two form  similar to eq. (\ref{Torsion-eq0}) as
\be\label{spinor-conn-eqn}
\alpha df\wdg\mathbf{\Omega}^{ab}+\frac{1}{\kappa^2}\theta^{a}\wdg \Theta^b-i*\theta^b\bar{\psi}\gamma_5\gamma^a\psi=0,
\ee 
after some algebra of the gamma matrices and making use of the definition $\gamma_5\equiv i\gamma^0\gamma^1\gamma^2\gamma^3$ ($i^2=-1$).

Although it is technically more involved, one can proceed in the same manner as the approximation scheme  introduced above to include the Dirac spinor field starting from the independent connection equations (\ref{spinor-conn-eqn}). Consequently, one can obtain  an approximate set of field equations in a reduced form with now involving scalar-spinor interactions as well.

\section{A de Sitter type cosmological solution }\label{sec6}
 
As an application of the results obtained in Section 4, in this section we consider an exact solution of the modified Gauss-Bonnet gravity models defined by Lagrangian (\ref{lag-def}) that we consider for both torsionless case and the case with torsion.  Most of the earlier works on the modified GB theory \cite{gasperini-giovannini,nojiri-odintsov-sasaki,nojiri-odintsov,Cognola,Koivisto-mota1,Koivisto-mota2,bamba,uddin,defelice,h-j-schmidt}   focused on the cosmological implications of this theory to discuss the late time acceleration of the universe and employed a power law scale factor. The late time expansion asymptotically approaches to a de-Sitter universe, and the issues related to the existence and stability of de-Sitter universe having an exponential scale factor for various modified GB-extended theories are studied, for example,  in \cite{cognola-elizalde-nojiri,capozziello-elizalde}. Similar to these works, for the sake of simplicity, we fix the spacetime metric as the de-Sitter metric  and obtain the field equations for arbitrary functions $f(\phi)$ and $V(\phi)$.  The maximally symmetric de-Sitter metric has the form
\begin{equation}\label{ds}
ds^2=-dt^2+e^{2Ht}(dx^2+dy^2+dz^2),
\end{equation}
with $H$ being a constant. We shall also assume that the scalar field depends only on the time coordinate as $\phi=\phi(t)$.
 
\subsection{Pseudo-Riemannian case}
Let us first consider the torsion-free theory. We can employ the following coframe one forms:
\begin{equation}
\theta^0=dt,\quad \theta^i=e^{Ht} dx^i.
\end{equation}
where $i=1,2,3$.
For this Riemannian case,
 the connection, curvature, dual curvature, Ricci and Einstein forms  and Ricci scalar read
\begin{eqnarray}
&&\omega^{0i}=H\theta^i, \omega^{ij}=0, \quad \Omega^{ab}=-\tilde{\Omega}^{ab} = H^2\theta^{ab},R^a=3H^2\theta^a,
 \\ &&
  G^a=-3H^2\theta^a,\quad R=12 H^2,\quad
 *T^{0}[\phi]=-\frac{1}{2}\dot{\phi}^2*\theta^0,\quad  *T^{i}[\phi]=\frac{1}{2}\dot{\phi}^2*\theta^i, 
\end{eqnarray}
respectively.
Using these expressions, the field equations (\ref{reduced-coframe-eqn}) and (\ref{scalar-fe}) become
\begin{eqnarray}
&&\frac{3H^2}{\kappa^2}-\frac{1}{2}\dot{\phi}^2-V(\phi)+6 \alpha H^3\,f'\,\dot{\phi}=0,\label{fe_Rim0}\\
&& \frac{3H^2}{\kappa^2}+\frac{1}{2}\dot{\phi}^2-V(\phi)+2 \alpha H^2\left( f'' \dot{\phi}^2+f'\ddot{\phi}+2Hf'\dot{\phi}
 \right)=0, \label{fe_Rim1}\\
&& \ddot{\phi}+3H \dot{\phi}+V'(\phi)-6 \alpha H^4 f' =0.\label{fe_sc}
\end{eqnarray}
In order to solve  eqs. (\ref{fe_Rim0},\ref{fe_Rim1},\ref{fe_sc}), one needs to specify the arbitrary function $f(\phi)$ and potential $V(\phi)$. We here consider the most simple case, i.e., $f(\phi)=\phi$ and $V=V_0=\mbox{constant}$ and able to  find the following solution:
\begin{eqnarray}\label{solrim}
\phi(t)=2 \alpha H^3\,t+b,\quad 
V_0=\frac{3H^2}{\kappa^2}+10\, \alpha^2 H^6 .\label{sol_rim}
\end{eqnarray}
This is an exact solution of a pseudo-Riemannian modified GB gravity model describing an exponential expansion of the universe and having maximally symmetric spacetime geometry given by the metric (\ref{ds}). Note that, de Sitter solutions of modified GB model with $\phi$ linear in $t$ as given above are known \cite{cognola-elizalde-nojiri,capozziello-elizalde}. Thus, it is interesting to explore whether such  simple solutions are also admitted  by the model with torsion as well.

  \subsection{The case with torsion}
  
  We apply the perturbative approach to obtain the field equations in this case. Since the first order case  yields the same equations, namely the equations given in (\ref{fe_Rim0}),  (\ref{fe_Rim1}) and (\ref{fe_sc}) with the same particular solution (\ref{sol_rim}),   we consider the field equations to the $O(\alpha^2)$ order. However,  in order to calculate field equations in this order, we first need  the expressions of the torsion and the contorsion  terms to the first order. Hence,  by inserting ${{\bf\Omega}}^{ab}=\Omega^{ab}$  into   (\ref{Torsion-eq1}), we obtain
  \begin{equation}
  \Theta^{a(1)}=\alpha \kappa^2 f'\dot{\phi} H^2 \, \theta^{a}\wedge\theta^0,
  \end{equation} 
  which yields 
  \begin{equation}
  K_{i0}=\alpha \kappa^2 f' \dot{\phi} H^2 \theta_i, \quad K_{ij}=0.
  \end{equation}
  We will also need the following terms,
  \begin{eqnarray}
&&DK_{0i}^{(1)}=\alpha \kappa^2 H^2 \left(f''\dot{\phi}^2+f' \ddot{\phi
} +H f' \dot{\phi} \right)\, \theta_0\wedge \theta_{i}  , \\
&& DK^{(1)}_{ij}= 2\alpha \kappa^2 H^3 f' \dot{\phi}\, \theta_{i}\wedge \theta_{j} .  
  \end{eqnarray} 
  Consequently, the calculation the field equations to the first order case leads exactly to the field equations  (\ref{fe_Rim0}),(\ref{fe_Rim1}) and 
(\ref{fe_sc}). Next, we can calculate  the field equations to second order in $\alpha$, hence one needs the torsion and contorsion  form expressions to second order in $\alpha$. To this end, by using (\ref{omega-1}) in (\ref{Torsion-eq}), we find
  \begin{equation}
\Theta^{a(2)}=2\alpha^2\, \kappa^4\,f'^2\,\dot{\phi}^2 \, H^3 \theta^a \wedge \theta^0  ,
  \end{equation}
  which yields
\begin{equation}\label{Kij}
K_{i0}^{(2)}=2\alpha^2\, \kappa^4\, f'^2 \dot{\phi}^2\, H^3\, \theta_i, \quad K_{ij}^{(2)}=0.
\end{equation}
From (\ref{Kij}), we have
\begin{eqnarray}
&&DK^{(2)}_{0i}=2\alpha^2\, \kappa^4 H^3 f'\dot{\phi} \left(2f''\dot{\phi}^2+2 f' \ddot{\phi}+f' \dot{\phi} H \right) \theta_{0}\wedge \theta_i,\label{DK0}\\
&&DK^{(2)}_{ij}=4 \alpha^2  \kappa^4\, H^4 f'^2 \dot{\phi}^2\,\theta_i\wedge \theta_j,\label{DKi}
\end{eqnarray}
and we also have
\begin{eqnarray}
&&K_{bd}^{(1)}\wedge K^{d(1)}_{\phantom{a}c}*\theta^{0bc}=6 \alpha^2 \kappa^4 H^4 f'^2\dot{\phi}^2  *\theta^0, \label{KvK0} \\
&& K_{bd}^{(1)}\wedge K^{d(1)}_{\phantom{a}c}*\theta^{ibc}=2 \alpha^2 \kappa^4 H^4 f'^2\dot{\phi}^2 *\theta^i, \label{KvKi}
\end{eqnarray}
 and by using the eqs. (\ref{DK0},\ref{DKi},\ref{KvK0},\ref{KvKi}), the metric field equations (\ref{fe2}) become
 \begin{eqnarray}
 &&\frac{3H^2}{\kappa^2}-\frac{1}{2}\dot{\phi}^2-V(\phi)+6\alpha H^3 f' \dot{\phi}+15 \alpha^2 \kappa^2 H^4 f'^2\dot{\phi}^2=0,\label{dstfe1}\\
 && \frac{3H^2}{\kappa^2}+\frac{1}{2}\dot{\phi}^2-V(\phi)+2\alpha H^2 \left( f'' \dot{\phi}^2+f' \ddot{\phi} +2 H f' \dot{\phi} \right) + 8 \alpha^2 \kappa^2 H^3 f'\dot{\phi} \left(  f'' \dot{\phi}^2+ f' \ddot{\phi}+ \frac{9}{8} H f' \dot{\phi}  \right)=0 \label{dstfe2}. 
 \end{eqnarray}
 Moreover, the scalar equation (\ref{scalara2}) becomes
 \begin{equation}
 \ddot{\phi}+3 H \dot{\phi} +V'(\phi)-6 \alpha f' H^4-6 \alpha^2 \kappa^2 H^4  \left(f' f'' \dot{\phi}^2+f'^2 \ddot{\phi}+3 H f'^2\dot{\phi}^2 \right)=0.\label{dstse}
\end{equation}  
The last terms proportional to $\alpha^2$ in the last three eqs. (\ref{dstfe1},\ref{dstfe2},\ref{dstse}) are due to the presence of  torsion which makes the Riemann-Cartan case different  from the corresponding pseudo-Riemannian case. In order to be able to solve these equations, the form of $f(\phi)$ and $V(\phi)$ must be specified.  Similarly to the previous subsection, we are able to find the solution with the same choice of $f$ and $V$, namely, $f(\phi)=\phi$ , $ V(\phi)=V_0=\mbox{constant}$ with 
\begin{equation} 
\phi(t)=\frac{2\alpha H^3}{1-6 \alpha^2\kappa^2 H^4}t+b, \quad V_0=\frac{3H^2-26 \alpha^2 H^6 \kappa^2+96 \alpha^4 \kappa^4 H^{10}}{(1-6 \alpha^2\kappa^2 H^4)^2 \kappa^2}.\label{solntor}
\end{equation}

Thus, we have seen that, the maximally symmetric de-Sitter metric given in (\ref{ds}) actually solves the modified Gauss-Bonnet gravity in four spacetime dimensions for both torsionless case as well as case with torsion up to second order correction terms included, for a particular nontrivial but simple choice of the scalar field $\phi$.
To compare the solutions we series expand (\ref{solntor}) in $\alpha$ as
\begin{equation}
\phi(t) \approx \left[2\alpha H^3+12 \alpha^3 \kappa^2 H^7+O(\alpha^5)\right] t+b, \quad V_0\approx \frac{3H^2}{\kappa^2}+10\alpha^2 H^6 +108 \alpha^4 \kappa^2 H^{10}+O(\alpha^6)
\end{equation}
and see that the above solution (\ref{solntor}) successfully include Riemannian one (\ref{sol_rim}) and the second term in $\phi$ and third term in $V_0$ are the leading order correction terms due to the presence of torsion in the solution.

 The solutions above for the modified  GB-modified gravity model  is important  for the viability of these theories/models and the compatibility  of such theories with the observations. This is because  the past inflationary  epoch described in GR by de-Sitter geometry in the presence of a cosmological constant. Also, the present accelerated expansion can be attributed  to the dominant dark energy asymptotically approaches to a de-Sitter space-time.  Note also that in the GR case, the relation of expansion rate $H$ with the cosmological constant $\Lambda$ is of the form $H \sim \sqrt{\Lambda}$, whereas the corresponding  relation with $V_0$ in the  modified GB model that in the pseudo-Riemannian  case and the case with torsion are different and they are explicitly given by Eqs. (\ref{solrim}) and (\ref{solntor}), respectively. The above solutions  are not in conflict with the results obtained prevously by  Gasperini and Giovanini \cite{gasperini-giovannini} because there is no potential term in their analysis.

\section{Concluding remarks}

In this paper a modified Gauss-Bonnet gravity is investigated  in four spacetime dimensions where the Gauss-Bonnet action term is coupled with a function of an arbitrary scalar field. First, by making use of the first order formalism, the Riemannian theory  following form the model is studied where the torsion-free condition is imposed via a Lagrange multiplier 2-form. After obtaining the Lagrange multiplier form, the field equations were presented in a compact form with the help of double-dual curvature form in a convenient form. In the pseudo-Riemannian case, the additional condition on the scalar function introduced by G\"urses \cite{gurses}  is considerably simplifies field equations and the use of double-dual curvature 2-form in expressing the coframe field equations elucidates the choice of the particular constraint (\ref{gurses-assumption}) on the scalar function.

 Previously,  Gasperini and Giovannini \cite{gasperini-giovannini}  derived the field equations in the pseudo-Rimenannian case by independent connection and coframe variational derivatives with respect to independent connection and coframe forms.  Slightly more rigorous treatment given above, with the help of a Lagrange multiplier imposing the torsion-free condition on a metric compatible connection 1-form, shows that the field equations governing the lagrangian $L_{EGB}$ in the Riemann-Cartan case can only approximately be identified with the corresponding equations of the pseudo-Riemannian case which also confirms their method to  derive the field equations in pseudo-Riemannian case. To discuss the effect of a nonvanishing torsion to some extent,
a perturbative  scheme is introduced to express the field equations in  powers of a weak constant $\alpha$. In the approximation scheme, the field equations to  first order turn out to be  identical to those of the pseudo-Riemannian case and the effect of nonvanishing terms arise  in the second order in powers of the coupling constant $\alpha$. We also note that the discussion above is confined to four spacetime dimensions and in the spacetime with dimensions $n>4$, there 
will be an additional 4-form term of the explicit form 
\be
\frac{\alpha}{4}{\bf\Omega}_{bc}\wedge {\bf \Omega}_{de}*\theta^{abcde}
\ee
in the metric (coframe) equations, in accordance with the results originally obtained by Gasperini and Giovannini.
This issue,  nicely pointed out by De Felice and Tsujikawa \cite{felice-tsujikawa}, also persists in modified GB gravity with the Lagrangian 
density of the form $f(\mathcal{G})$. To be mathematically more precise, in spacetime dimensions $n>4$, 
the 4-forms $*E^a$ obtained by a coframe variational derivative given in eq. (\ref{mGB-equations}), acquire an additional 4-form of the explicit form
\be\label{extra-term-conc}
\frac{1}{4}f'(\mathcal{G})\Omega_{bc}\wdg\Omega_{de}\wdg *\theta^{abcde}
=
-f'(\mathcal{G})
\left(
RR^{a}_{\fant{a}b}
-
2{R}^{ac}{R}_{cb}
+
2{R}_{cd} {R}^{cad}_{\fant{aaa}b}
+
{R}^{acde}{R}_{bcde}
-
\delta^{a}_{b}\mathcal{G}
\right)
*\theta^b
\ee
which is sometimes overlooked in the literature \cite{davis,li-barrow-mota}. It is evident from the expression on the left hand side of the Eq.
(\ref{extra-term-conc}) that a basis 5-form is not defined in four spacetime dimensions hence such a term is absent in four spacetime dimensions. 
It is less obvious, however, to draw the same conclusion from the equivalent (coordinate) expression on the right hand side of Eq. (\ref{extra-term-conc}) in terms of the components of the curvature tensor.

As an application of the results mentioned above,  we also considered the maximally symmetric  spacetime metric in a particular form, calculate the field equations and solve them in both of the  theories. This solution has a nontrivial, albeit simple, scalar field which has a  linear dependence on the time coordinate. It is interesting to see that the modified Gauss-Bonnet model we have considered in this work admits exponentially expanding de-Sitter universe as a solution for both the pseudo-Riemannian case, discussed by previously in \cite{cognola-elizalde-nojiri,capozziello-elizalde} and  also in the non-Riemannian case with torsion as well.

One future direction for  the current work is to consider the effects of torsion generated by a nonminimally coupled scalar field towards a cosmological application in the presence of matter sources in this non-Riemannian modified GB extended model.  Likewise, it is possible to extend the brief discussion of  the minimally coupled spinorial matter couplings presented in Section 5.

\section*{Appendix: Calculational  details for the derivation of $O(\alpha^2)$-order terms} \label{ap1} 

We present the explicit expression for the equation (\ref{H2}) which are of  $O(\alpha^2)$ correction terms due to the presence of a dynamic torsion.

Let us start with the first term in this equation. For calculational convenience, with the help of antisymmetrization brackets $A_{[ab]}=1/2\left( A_{ab}-A_{ba}\right)$,  we can express first order contorsion equation (\ref{K1}) in a compact form as
\begin{equation}
K_{ab}^{(1)}= 2\kappa^2\alpha \,  i_{c}( df)\left[  i^c \Omega_{ab} +2R_{[a}\, \delta_{b]}^c  + \theta_{[a}  G^{\phantom{a}c}_{b]}\right].
\end{equation} 
Then, after a straightforward but somewhat tedious exterior algebra manipulations, the first term turns out to have the form 
\begin{eqnarray}
&&K_{bd}^{(1)}\wedge K^{d(1)}_{\fant{dc}c}*\theta^{bca}=\nonumber\\
&=&4 \kappa^4 \alpha^2 (i_{e} df) (i_{g} df)\bigg\{ \left[ i^e \Omega_{bd} +2R_{[b}\, \delta_{d]}^e    + \theta_{[b}  G^{\phantom{a}e}_{d]}\right]  
\wedge 
\left[  i^g \Omega^{dc} +2R^{[d}\, \eta^{c]g}    + \theta^{[d}  G^{c]g} \right] \bigg\} *\theta^{b\phantom{c}a}_{\phantom{b}c}
\nonumber\\
&=&4 \kappa^4 \alpha^2 (i_{e} df) (i_{g} df)*\bigg\{  
\bigg[
 -R_{bd\phantom{e}c}^{\phantom{kl}e} R^{dcgb}   + 2 R_{b\phantom{de}c}^{\phantom{b}ge} R^{bc}        \, 
\nonumber
 + R_{bc}\, R^{bc}\, \eta^{eg} - \frac{7}{2}R_{b}^{\phantom{k}e} R^{bg}  +  \frac{5}{2} R^{ge}R   -\frac{5}{8} \eta^{eg} R^2    
\bigg]\theta^a    
\\
&&+ \bigg[4 R^{ce} R_{cb}^{\phantom{dc}ga} +  2 R_{cd}^{\phantom{kl}ea} R^{d\phantom{c}gc}_{\phantom{a}b}  + 2 R_{b}^{\phantom{b}gea} R - 2 R_{b\phantom{ka}c}^{\phantom{b}ge} R^{ac} 
 + \left( R_{b\phantom{ae}}^{\phantom{b}aec} 
   +  R_{b}^{\phantom{b}cea} \right) G_c^{\phantom{c}g}  \bigg]\theta^{b} \nonumber
  \\
 &&
   -3G^{ge} R^{a}   +\left(\eta^{ea}   G_b^{\phantom{c}g}  -2 R_{b}^{\phantom{b}gea}   -2 \eta^{eg}  R^a_{\phantom{a}b}\,   
     \right)   R^b         
 +2 G_{b}^{\phantom{b}g} R^{ab} \theta^{e} 
     +G^{ae} G^{g}  +2 R^{ag}\, R^{e}\,  
\bigg\}.  
\end{eqnarray}
To find the second term in $*H^{a(2)}$ we first calculate $K_{bc}^{(2)}\wedge *\theta^{abc}$ and subsequently calculate its covariant exterior derivative 
with $K_{bc}^{(2)}$ is given in (\ref{kontorsion-eq-2}). The resulting expression can be written in the form
\begin{eqnarray}
D K_{bc}^{(2)}\wedge  * \theta^{abc}
&=&   2\kappa^2\alpha\, D  * \bigg\{
  (i_{d} df)\, i^b \left[D K^{d(1)}_{\phantom{b}b} \right] \wedge \theta^{a}- i_{d}\left[df\wedge D K^{da(1)}  \right] 
\nonumber \\ 
 && 
+3\left( \theta^b \wedge i^a -\theta^a \wedge i^b \right) \left[ i_d \left( df\wedge  D K_{\phantom{b}b}^{d(1)}  \right) \right]
   \bigg\}  \quad,
\end{eqnarray}
with
\begin{eqnarray}\label{DK1}
D K_{ab}^{(1)}&=& 2\kappa^2\alpha \,D\left\{ (i_{c} df)\left[  i^c\Omega_{ab} +2 R_{[a}\, \delta_{b]}^c  + \theta_{[a} G^{\phantom{a}c}_{b]}\right]\right\}.
\end{eqnarray}

\section*{Acknowledgments}
We would like to thank the anonymous referee for the constructive comments and criticism  that helped to improve the manuscript.

\end{document}